# Wavelength flattened directional couplers: a geometrical approach

**Matteo Cherchi**

Pirelli Labs – Optical Innovation, viale Sarca 222, 20126 Milan, Italy

Abstract: A new approach to design wavelength insensitive optical power splitter is presented. First coupled-mode theory is cast in operatorial form. This allows to solve the equivalent of coupled differential equations as simple limits. The operators are then represented on a generalized Poincaré sphere, and the resulting graphical tool is applied to different structures, giving a clear interpretation of previous results in literature as well as hints on how to find improved solutions.

# Introduction

The need for wide band optical communication demands optical devices able to cover the whole band of interest. A very basic element of a typical device is, of course, the power splitter.

Power exchange between optical waveguides is usually obtained via directional couplers[1,2,3]. The simplest splitting device is the synchronous coupler, obtained keeping two identical waveguides close together over a given length. This coupling mechanism is based on the difference in propagation constants of the coupler super-modes. On the other hand this means that it can strongly depend on the wavelength.

Many proposals have been presented to overcome this problem. For example coupling different waveguides[4,5,6] (asynchronous coupler), tapering the coupled waveguides[7] (tapered coupler), or implementing interferometric structures[8,9,10,11] (Mach-Zehnder coupler). Takagi et al.[4] fabricated an asynchronous coupler with a coupling ratio of (50±5)% over a wavelength range of 400nm, as well as a tapered coupler[7] achieving (50±5)% over 500nm. Jinguji et al.[11] built a Mach Zehnder coupler that achieves a splitting ratio of (50±1.9)% over 400nm, and with a similar structure Gonthier et al.[9] obtained (50±2.5)% over 300nm.

In this paper we will focus on wavelength sensitivity of 50/50 splitters, but a similar approach applies to all splitting ratios as well as to changes of parameters other than wavelength.

# Unitary Transformations

We will assume a power splitter as a lossless device made of two waveguides, one beside the other over a length L along the propagation axis *z* (see Fig. 1). In general their cross-sections, their distance and their refractive index profiles may vary along *z*. We will also assume that, for any fixed $z_0$, the two eigenmodes of the single waveguides cross-section, $E_j[z_0] \equiv E_j(x, y, z_0)$, ($j=1,2$) can be considered as a basis[12] for the field at that point, so that it is always possible to write:

$$E[z] \equiv a_1[z] \, E_1[z] + a_2[z] \, E_2[z]$$

where $a_j[z]$ are complex numbers so that $|a_1[z]|^2 + |a_2[z]|^2 = 1$ and $E_j[z]$ are normalized so that $|a_j|^2$ represent the power fractions in each of the waveguides. We will regard $E[z]$ as a scalar quantity, since we will consider only singly copolarized modes of the structure. With these assumptions we can always represent the generic state of the system through the complex vector

$$\mathbf{u}[z] \equiv \begin{pmatrix} a_1[z] \\ a_2[z] \end{pmatrix}. \tag{1}$$

Its evolution can be described by an unitary 2x2 matrix or, up to an overall phase[13], by an element $\mathbf{U}$ of SU(2) (i.e. an unitary matrix so that $\det \mathbf{U} = 1$). From its definition, the most general form for $\mathbf{U}$ is[14]:

$$\begin{pmatrix} \exp(-i\theta/2)\cos\phi & \exp(i\xi)\exp(-i\theta/2)\sin\phi \\ -\exp(-i\xi)\exp(i\theta/2)\sin\phi & \exp(i\theta/2)\cos\phi \end{pmatrix}. \tag{2}$$

If optical power is launched in just one of the two waveguides, the splitting ratio (defined as the power fraction transferred to the other waveguide) does not depend on which waveguide is chosen and it is clearly given by $\sin^2\phi$.

## Global View

Eq. 2 describes a power splitter as a black box which propagates the input state from $z=0$ to $z=L$, i.e. so that $\mathbf{u}[L] = \mathbf{U}[L,0]\,\mathbf{u}[0]$. It can be cast in the form $\mathbf{U} = \mathbf{V}(\theta)\,\mathbf{W}(\phi,\xi)$ where:

$$\mathbf{V}(\theta) \equiv \begin{pmatrix} \exp(-i\theta/2) & 0 \\ 0 & \exp(i\theta/2) \end{pmatrix} \quad (3)$$

$$\mathbf{W}(\phi,\xi) \equiv \begin{pmatrix} \cos\phi & \exp(i\xi)\sin\phi \\ -\exp(-i\xi)\sin\phi & \cos\phi \end{pmatrix} \quad (4)$$

which physically means that any unitary transformation can be decomposed in the product of a $\theta$-phase shift between the two branches and a coupling in which the field transferred from the upper (or lower) branch acquires a $\xi$ (or $\pi-\xi$)-phase shift.

## Local View

If we look locally at the whole transformation $\mathbf{U}[L,0]$, we can decompose it in an infinite number of infinitesimal unitary transformations $\mathbf{U}_n \equiv \mathbf{U}[z_{n+1}, z_n]$:

$$\mathbf{U}[L,0] = \lim_{N\to\infty} \overleftarrow{\prod_{n=0}^{N-1}} \mathbf{U}_n \quad (5)$$

where $\overleftarrow{\prod}$ means matrix product on the left, and we have defined $\Delta z \equiv L/N$, $z_n \equiv n\Delta z$ (from now on $f_n \equiv f(z_n) \;\forall f(z)$).

From the physical interpretation of Eqs. 3,4 we can think of every small section of the splitter as a composition of a local phase shifter $\mathbf{V}(\theta_n)$ with $\theta_n \equiv \Delta\beta_n \Delta z$ (where $\Delta\beta_n \equiv \beta_2[z_n] - \beta_1[z_n]$ is the local difference in propagation constants of the two branches) and a local coupler $\mathbf{W}(\phi_n, \xi_n)$ with $\phi_n = \kappa_n \Delta z$ (where $\kappa_n$ is the local coupling coefficient between the two waveguides and $\xi_n$ is the phase acquired locally, passing from the first to the second branch). To 1$^{st}$ order in $\Delta z$ can rewrite

$$\mathbf{U}_n \equiv \mathbf{I} + \boldsymbol{\tau}_n \Delta z$$

where $\mathbf{I}$ is the identity matrix and

$$\boldsymbol{\tau}_n \equiv \begin{pmatrix} -\tfrac{i}{2}\Delta\beta_n & \kappa_n \exp(i\xi_n) \\ -\kappa_n \exp(-i\xi_n) & \tfrac{i}{2}\Delta\beta_n \end{pmatrix}. \tag{6}$$

$\mathbf{U}_n$ is the infinitesimal propagator from $z_n$ to $z_{n+1}$, i.e. the matrix so that

$$\mathbf{u}_{n+1} = \mathbf{U}_n \mathbf{u}_n. \tag{7}$$

Notice that Eq. 7 is nothing but a compact form for writing the standard[1,2,3] system of coupled differential equations in the coefficients $a_1[z]$, $a_2[z]$ of Eq. 1.

## Special Cases

In some special cases Eq. 5 can be cast in closed form:

1) Null coupling.

Setting $\kappa = 0$ gives $\mathbf{U}[L,0] = \mathbf{V}(\Theta)$ with $\Theta \equiv \int_0^L \Delta\beta \, dz$. This result describes a 'pure' phase shifter and it is easily shown by noticing that $\mathbf{V}(\theta_1)\,\mathbf{V}(\theta_2) = \mathbf{V}(\theta_1 + \theta_2)$.

If we define $\langle\beta[z]\rangle \equiv (\beta_1[z] + \beta_2[z])/2$, the neglected overall phase[13] is $\exp(i\eta)$, with

$$\eta \equiv \int_0^L \langle\beta[z]\rangle \, dz.$$

2) Null phase shift.

Imposing $\beta_1 = \beta_2 = \beta$ and $d\xi/dz = 0$ gives $\mathbf{U}[L,0] = \mathbf{W}(\Phi,\xi)$ with $\Phi \equiv \int_0^L \kappa \, dz$.

This describes a 'pure' coupler and, as before, it is a consequence of the property $\mathbf{W}(\phi_1,\xi)\,\mathbf{W}(\phi_2,\xi) = \mathbf{W}(\phi_1+\phi_2,\xi)$.

In this case the neglected overall phase amounts to $\exp(i\eta)$, with $\eta \equiv \int_0^L \beta\,\mathrm{d}z$.

Actually the only coupling mechanism known to the author is reciprocal, i.e. time reversal. It can be shown[2] that the time reversal combined to the unitarity condition implies $\mathbf{U}^\dagger = \mathbf{U}^*$, that in Eq. 2 requires $\xi = \pi/2$. Physically this means that the phase acquired coupling from the first to the second branch must be equal to the phase acquired in the reverse process. Nevertheless we will allow for generic $\xi$ values, so that our formalism holds for any unitary system (e.g. for polarization states, where the Faraday rotation is non-reciprocal).

3) Constant coupling and phase shift (asynchronous coupler).

Suppose $\mathrm{d}\Delta\beta/\mathrm{d}z = 0$, $\mathrm{d}\kappa/\mathrm{d}z = 0$ and $\mathrm{d}\xi/\mathrm{d}z = 0$. Eq. 5 becomes:

$$\mathbf{U}[L,0] \equiv \overline{\mathbf{U}}[L,0] = \lim_{N\to\infty}\left(\mathbf{I}+\frac{L}{N}\boldsymbol{\tau}\right)^N = \lim_{N\to\infty}\sum_{k=0}^{N}\binom{N}{k}\left(\frac{L}{N}\boldsymbol{\tau}\right)^k = \lim_{N\to\infty}\sum_{k=0}^{N}\frac{(L\boldsymbol{\tau})^k}{k!} = \exp(L\boldsymbol{\tau}).$$

where, from now on, the overbar will denote the solution with constant coupling and phase shift. Let's now introduce the hermitian matrix $\boldsymbol{\upsilon} \equiv \boldsymbol{\tau}/(i\mu)$, where $\mu^2 \equiv \det\boldsymbol{\tau} = (\Delta\beta/2)^2 + \kappa^2$. It is easily verified that $\boldsymbol{\upsilon}^{2n} \equiv (-\mathbf{I})^n$. So the sum splits in even and odd terms giving:

$$\overline{\mathbf{U}}[L] = \cos(\mu L)\,\mathbf{I} + i\sin(\mu L)\,\boldsymbol{\upsilon}. \qquad (8)$$

The matrix $\boldsymbol{\upsilon}$ is the so called infinitesimal generator[14,15] of $\overline{\mathbf{U}}$.

In this case the neglected overall phase is clearly $\exp(i\langle\beta\rangle L)$, where

$$\langle \beta \rangle \equiv (\beta_1 + \beta_2)/2. \tag{9}$$

If we define $\cos\gamma \equiv -\Delta\beta/(2\mu)$ (which implies $\sin\gamma = \kappa/\mu$), Eq. 8 becomes

$$\overline{\mathbf{U}}[L] \equiv \overline{\mathbf{U}}(\mu L, \xi, \gamma) = \begin{pmatrix} \cos(\mu L) + i\sin(\mu L)\cos\gamma & \exp(i\xi)\sin(\mu L)\sin\gamma \\ -\exp(-i\xi)\sin(\mu L)\sin\gamma & \cos(\mu L) - i\sin(\mu L)\cos\gamma \end{pmatrix} \tag{10}$$

Since we are neglecting overall phases

$$\overline{\mathbf{U}}(\mu L + \pi, \xi, \gamma) = -\overline{\mathbf{U}}(\mu L, \xi, \gamma) \cong \overline{\mathbf{U}}(\mu L, \xi, \gamma) \tag{11}$$

i.e. $\overline{\mathbf{U}}(\mu L, \xi, \gamma)$ is π-periodic in $\mu L$.

Notice that in solving for the propagator matrix $\mathbf{U}[L,0]$ (Eq. 5), we don't need to impose any boundary conditions, as for usual differential equations, because they are implicit in the input state $\mathbf{u}[0]$ on which $\mathbf{U}[L,0]$ operates.

## *Eigenstates*

In order to interpret the last result, it is useful to determine the eigenstates of $\overline{\mathbf{U}}$. From the determinantal equation the eigenvalues are found to be:

$$\lambda_\pm = \exp(\pm i\mu L) \tag{12}$$

and, correspondingly, the eigenvectors can be cast in the form

$$\mathbf{u}_+ = \begin{pmatrix} \exp(-\tfrac{i}{2}(\pi/2 - \xi))\cos(\gamma/2) \\ \exp(\tfrac{i}{2}(\pi/2 - \xi))\sin(\gamma/2) \end{pmatrix}$$

$$\mathbf{u}_- = \begin{pmatrix} -\exp(-\tfrac{i}{2}(\pi/2-\xi))\sin(\gamma/2) \\ \exp(\tfrac{i}{2}(\pi/2-\xi))\cos(\gamma/2) \end{pmatrix}. \tag{13}$$

From Eq. 3,4, defining $\mathbf{R}(\gamma,\xi) \equiv \mathbf{V}(\pi/2-\xi)\,\mathbf{W}(\gamma/2,0)$, we can also rewrite

$$\mathbf{u}_{+,-} = \mathbf{R}(\gamma,\xi)\,\mathbf{u}_{1,2}$$

$$\overline{\mathbf{U}}(\mu L,\xi,\gamma) = \mathbf{R}(\gamma,\xi)\,\mathbf{V}(-2\mu L)\,\mathbf{R}^{-1}(\gamma,\xi). \tag{14}$$

Physically Eq. 12 means that the difference between the propagation constants of the two eigenstates is $\Delta\beta_{\mp} \equiv \beta_- - \beta_+ = -2\mu$. On the other hand Eq. 14 means that $\overline{\mathbf{U}}(\mu L,\xi,\gamma)$ is nothing but a $\mathbf{V}(\Delta\beta_{\mp} L)$ diagonal transformation (i.e. a transformation represented in the basis of its eigenstates) represented in the rotated basis $\{\mathbf{R}^{-1}(\gamma,\xi)\mathbf{u}_+, \mathbf{R}^{-1}(\gamma,\xi)\mathbf{u}_-\}$.

Looking at $\overline{\mathbf{U}}$ globally, we can also rewrite the neglected overall phase as[13] $\exp(i\langle\beta_{\mp}\rangle L)$, where $\langle\beta_{\mp}\rangle \equiv (\beta_+ + \beta_-)/2$. From Eq. 9 (which was obtained looking at the local $\mathbf{U}_n$) we get $\langle\beta_{\mp}\rangle = \langle\beta\rangle$, which implies $\beta_{\mp} = \langle\beta\rangle \mp \mu$.

Notice that, for $\kappa = 0$, the eigenstates of $\overline{\mathbf{U}}$ do not depend on $\Delta\beta$, so they must also be the eigenstates of $\mathbf{V}(\Theta)$ (the first special case discussed earlier). Similarly, for $\Delta\beta = 0$ they do not depend on $\kappa$, so they must also be the eigenstates of $\mathbf{W}(\Phi,\xi)$.

## Generalized Poincaré Sphere

All the results obtained so far are better understood introducing a convenient geometrical representation. It is well known[15,16] that SU(2) transformations may be mapped into SO(3) transformations through a homomorphism. This means that all the transformation we have analyzed before can be represented as rotations on a spherical surface, analogous of the Poincaré sphere[17] for polarization states. In Fig. 2 are displayed all the intersection of the $S_1$, $S_2$, $S_3$ axes with the sphere. They represent the single waveguides modes $E_1$, $E_2$ and their linear combinations $E_{S,A} \equiv \frac{1}{\sqrt{2}}(E_1 \pm E_2)$ and $E_{R,L} \equiv \frac{1}{\sqrt{2}}(E_1 \pm i E_2)$. Also is plotted the generic normalized mode $P \equiv a_1 E_1 + a_2 E_2 \equiv \cos\alpha\, E_1 + \exp(i\theta)\sin\alpha\, E_2$, with Stokes parameters[17]

$$\begin{aligned} S_0 &\equiv |a_1|^2 + |a_2|^2 = 1 \\ S_1 &\equiv |a_1|^2 - |a_2|^2 = \cos 2\alpha \\ S_2 &\equiv a_1 a_2^* + a_1^* a_2 = \sin 2\alpha \cos\theta \\ S_3 &\equiv i(a_1 a_2^* - a_1^* a_2) = \sin 2\alpha \sin\theta \end{aligned}.$$

In Fig. 3 we have plotted the eigenmodes $E_{+,-}$ (corresponding to the eigenstates $\mathbf{u}_{+,-}$ of Eq. 13). It is clear that the generic transformation $\overline{\mathbf{U}}$ (Eq. 8), corresponds, on the sphere, to a rotation about the axis of its eigenstates $\overline{E_+ E_-} \equiv S(\gamma, \pi/2 - \xi)$ (which is the rotated of the axis $S_1$ by an angle $\gamma$ about $S_3$ and then by an angle $\pi/2 - \xi$ about $S_1$). This means that a generic input state $P$ will be rotated by an angle $-2\mu L$ (remember Eq. 14) on the circle of revolution about $S(\gamma, \pi/2 - \xi)$ passing through $P$. The three special cases discussed earlier may be represented as rotations on the sphere:

1) $\kappa = 0$ implies $E_{+,-} = E_{1,2}$ and $\mathbf{U} = \mathbf{V}(\Theta)$.

Physically it is clear that $E_{1,2}$ are the eigenmodes of the phase shifter. On the sphere a generic input state $P$ will be rotated by an angle $\Theta$ on the circle of revolution about $S_1$ (Fig. 4) passing through $P$. In the special case of constant phase shift will be $\Theta = \Delta\beta L$.

Notice that the circles of revolution about $S_1$ represent the loci of constant power splitting, which we will call *isodias* ($\iota\sigma o$ = equal, $\delta\iota\alpha$ = split in two parts).

2) $\Delta\beta = 0$ and $\xi = \pi/2$ means $E_{+,-} = E_{S,A}$ and $\mathbf{U} = \mathbf{W}(\Phi, \pi/2)$.

Physically it is clear that the symmetric and antisymmetric superpositions of the single waveguide modes are the eigenmodes of the synchronous coupler. On the sphere a generic input state $P$ will be rotated by an angle $2\Phi$ on the circle of revolution about $S_2$ (Fig. 5) passing through $P$. In the special case of constant coupling will be $-2\Phi = -2\kappa L = \Delta\beta_{AS} L$, where we have defined $\Delta\beta_{AS} \equiv \beta_A - \beta_S$.

If one allows for generic $\xi$ ($\mathbf{U} = \mathbf{W}(\Phi, \xi)$), it is easily seen that the loci of constant $(\pi/2 - \xi)$ phase shift are the semicircles with diameter on $S_1$, which we will call *isophases* (notice that isodias and isophases can be regarded as a parallels and meridians for the sphere, with poles on $E_1$ and $E_2$).

3) $\Delta\beta \neq 0$ and $\xi = \pi/2$ means $\mathbf{u}_{+,-} = \mathbf{W}(\gamma/2, 0)\mathbf{u}_{1,2}$ and $\mathbf{U} \equiv \overline{\mathbf{U}}(\mu L, \pi/2, \gamma)$.

In the limit of validity of our approximation[12], the eigenmodes of an asynchronous coupler are similar to $E_{S,A}$, but with unbalanced power in the waveguides. On the sphere a generic input state $P$ will be rotated by an angle $-2\mu L = \Delta\beta_{\mp} L$ on the circle of revolution about $S(\gamma, 0)$ (Fig. 6) passing through $P$.

## 50/50 Splitters

We will now analyze the wavelength dependence of some 50/50 splitters. Under a change $d\lambda$ in the wavelength $\lambda$, to any $f(\lambda)$ will correspond a relative change $df/f$ (relative changes are easier to calculate when working with products and divisions resembling the rules of relative error propagation).

On the sphere a 50/50 splitter is any transformation which, starting from $E_1$ (or, equivalently, from $E_2$), reaches any point on the isodia $\Gamma$ with diameter on $S_2$.

1) The straightest way to do so is by a synchronous coupler, i.e. by a $\mathbf{W}(\pi/4,\pi/2)$ transformation. Power transfer is given by $P_2 = \tfrac{1}{2}(1-\cos(\Delta\beta_{AS}L))$, and varies as:

$$dP_2 = \frac{1}{2}\sin(\Delta\beta_{AS}L)\,L\,d\Delta\beta_{AS} = \frac{1}{2}\Delta\beta_{AS}L\,\sin(\Delta\beta_{AS}L)\left(\frac{d\Delta n_{AS}}{\Delta n_{AS}} - \frac{d\lambda}{\lambda}\right) \quad (15)$$

where $\Delta n_{AS} \equiv n_A - n_S$ is the difference between the effective indexes of the symmetric and antisymmetric modes. In our case

$$dP_2 = \frac{1}{2}L\,d\Delta\beta_{AS} = \frac{\pi}{4}\left(\frac{d\Delta n_{AS}}{\Delta n_{AS}} - \frac{d\lambda}{\lambda}\right).$$

So the only way to minimize the wavelength sensitivity is by a convenient choice of the waveguides and of their distance[18].

On the sphere (Fig. 7) we observe that power splitting is measured by the $S_1$ parameter (that, by definition, is the difference between optical power in the two waveguides). Since a synchronous coupler approaches the isodia $\Gamma$ parallel to $S_1$, a change in the angle $L\Delta\beta_{AS}$ translates immediately into a power change.

2) An alternative approach could be using an asynchronous coupler[4,5,6]. Power transfer in this case is given by $P_2 = P_0 F$, where $P_0 \equiv \sin^2\gamma = 1 - (\Delta\beta/\Delta\beta_\mp)^2$ is the maximum transferable power and $F \equiv \tfrac{1}{2}(1-\cos(\Delta\beta_\mp L))$ is the power oscillation along propagation. A $\lambda$ change will now give $dP_2/P_2 = dP_0/P_0 + dF/F$ where

$$dP_0 = 2(P_0 - 1)\left(\frac{d\Delta n}{\Delta n} - \frac{d\Delta n_\mp}{\Delta n_\mp}\right) \equiv 2(P_0 - 1)\delta \quad (16)$$

$$dF = \frac{1}{2}\sin(\Delta\beta_\mp L)\, L\, d\Delta\beta_\mp = \frac{1}{2}\Delta\beta_\mp L\, \sin(\Delta\beta_\mp L)\left(\frac{d\Delta n_\mp}{\Delta n_\mp} - \frac{d\lambda}{\lambda}\right) \qquad (17)$$

having defined $\Delta n \equiv n_2 - n_1$ the difference in effective index of the single waveguides modes.

In particular when $\Delta\beta_\mp L = k\pi$, the oscillating contribution vanishes and we can get a 50/50 splitter if $\gamma = \pi/4$. Regarding the $dP_0$ contribution we notice that it is null for $\gamma = m\pi/2$ and it doesn't feature any $d\lambda$ term. So, to get an insensitive coupler, a convenient choice of the coupler parameters must made $\delta$ become negligible.

On the sphere (Fig. 8) $dP_0$ is due to a $\gamma$ change, i.e. a change of the rotation axis, while $dF$ is related to a $\Delta\beta_\mp$ change, i.e. a change of the rotation angle. It is clear that this asynchronous coupler approaches the isodia $\Gamma$ perpendicular to $S_1$ direction, so that, if the rotation axis doesn't change, power splitting it is invariant under $\Delta\beta_\mp$ changes.

This approach can be generalized cascading $N$ asynchronous couplers so that $\Delta\beta_\mp^{(i)} L^{(i)} = k\pi \ \forall i$ and $\sum_{i=1}^{N}(-)^{i+N}\gamma_i = \pi/4$. In Fig. 9 it is shown the case $\gamma_1 = \pi/8$ and $\gamma_2 = 3\pi/8$. Under the hypothesis that $d\gamma_1 = d\gamma_2$, we expect this configuration to solve the problem of rotation axis changes as shown in Fig. 10 (in general it will be true for any choice of $\{\gamma_i\}$ so that $\sum_{i=1}^{N}(-)^{i+N} d\gamma_i = 0$).

Another approach may be, instead of setting to zero each single contribution, to find a combination of $\gamma$ and $\Delta\beta_\mp L$ reaching a point on $\Gamma$ that make $dP_0/P_0$ and $dF/F$ cancel each other out.

These simple examples show how using a pictorial view can help not only to interpret well known results, but also to find better solutions.

3) Having understood the working principle of the asynchronous coupler, it seems that even better insensitivity could be obtained if we approached Γ descending from points closer to $E_R$. This could be done with a structure that starts as a pure synchronous coupler and, along propagation, becomes an almost pure phase shifter. This is the tapered coupler[7], schematically shown in Fig. 11. The trajectory on the sphere (Fig. 12) can be seen as a composition of small rotations about different axes $S(\gamma(z),0)$, with $\gamma(0) = \pi/2$ and $\gamma(L) \approx 0$. Notice that this structure is tolerant at the beginning (see Eq. 15) and at the end, but a dilatation (contraction) in the middle part of the trajectory (see Eq. 16 17), could shift the ending part of the trajectory on an isodia different by Γ. So, from a pictorial point of view, it is not clear whether a tapered coupler may be better than an asynchronous coupler. The answer can come only from a numerical study, and will depend on the parameters of the specific structure under investigation.

4) A completely different approach is based on an interferometric scheme[8,9,10,11]. Consider a synchronous directional coupler so that (at a certain working wavelength $\lambda_0$) $\phi_0 = \kappa_0 L = \pi/2$ (100% power transfer), cascaded with a $\theta$–phase shifter and another directional coupler, identical to the first one, but half the length (50% power transfer). At a generic $\lambda$ will be $\phi = \phi_0 + \Delta\phi$ and the system will be described by the matrix

$$\mathbf{M} = \mathbf{W}(\tfrac{1}{2}(\pi/2+\Delta\phi),\pi/2) \, \mathbf{V}(\theta) \, \mathbf{W}(\pi/2+\Delta\phi,\pi/2).$$

Of course when $\Delta\phi = 0$ this is perfectly equivalent to a $3\pi/4$ 50% splitter disregarding the $\theta$–value.

Assigned a (non infinitesimal) fixed value to $\Delta\phi$, our aim is to determine a corresponding $\theta$–value that still gives 50/50 power splitting. This means requiring $|M_{11}|^2 = 1/2$, which gives:

$$\cos\theta = \frac{t(t-1)-1/2}{1-t^2} \tag{18}$$

where $t \equiv \sin\Delta\phi$ and the condition $|\cos\theta| \leq 1$ implies $t \leq 1/2$. For small $\Delta\phi$ we have $\cos\theta \approx -1/2 - t \rightarrow -1/2$ or $\theta \rightarrow \pm 2\pi/3 + 2k\pi$.

This result is easily understood on the sphere (Fig. 13). Since we have a $\Delta\phi$ angular shift in the first coupler and a $\Delta\phi/2$ angular shift in the second coupler, a $2\pi/3$ rotation about $S_1$, which goes from $\phi_0 + \Delta\phi$ to $\phi_0 - \Delta\phi/2$, will compensate, at once, the shifts of both couplers (being the change of the circle representing the second coupler a 2$^{nd}$ order effect).

Notice that, in general, also the phase shifter will be wavelength dependent. In the case of a 'concentrated' phase shifter, made of two identical waveguides of different length and effective index $n$, it will be $d\theta/\theta = dn/n - d\lambda/\lambda$. In the case of a 'distributed' phase shifter, made of two different waveguides of the same length and a difference $\Delta n$ in their effective indexes, it will be $d\theta/\theta = d\Delta n/\Delta n - d\lambda/\lambda$. So it may be convenient to set the working point of the couplers and the working point of the phase shifter at different wavelengths in the desired band.

This example shows the power of the geometrical representation, that becomes apparent especially when dealing with interferometric schemes.

## Conclusions

We have cast coupled-mode theory in operatorial form. This formalism allows to solve the equivalent of the usual differential equations as simple limits. Furthermore the homomorphism between the SU(2) group and the SO(3) group allows to represent all these transformations on a generalized Poincaré sphere, which is found to be a powerful tool to design and understand tolerant 2x2 devices, once determined all parameter dependences.

## Captions

1. Schematic of a splitter

2. The Generalized Poincaré sphere

3. The eigenstates of $\overline{\mathbf{U}}(\mu L, \xi, \gamma)$

4. Action of a phase shifter

5. Action of a synchronous coupler

6. Action of an asynchronous coupler

7. A 50/50 synchronous coupler

8. A 50/50 asynchronous coupler

9. A 50/50 double asynchronous coupler

10. Insensitivity of the configuration of Fig. 9

11. Schematic of a tapered coupler

12. A 50/50 tapered coupler

13. The 50/50 Mach-Zehnder coupler

**Figures**

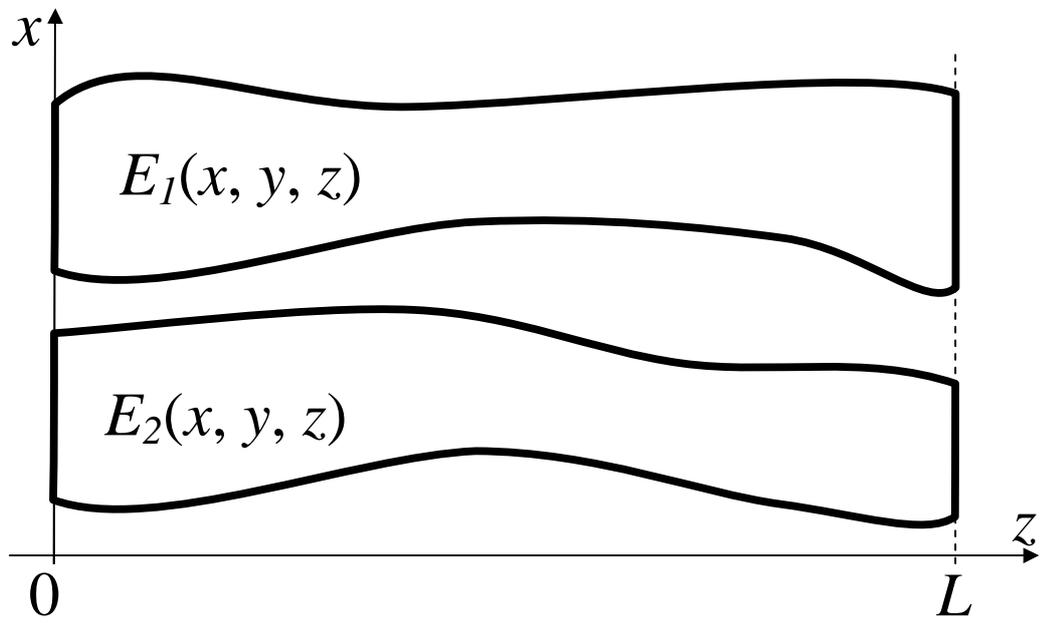

Fig. 1: Schematic of a splitter.

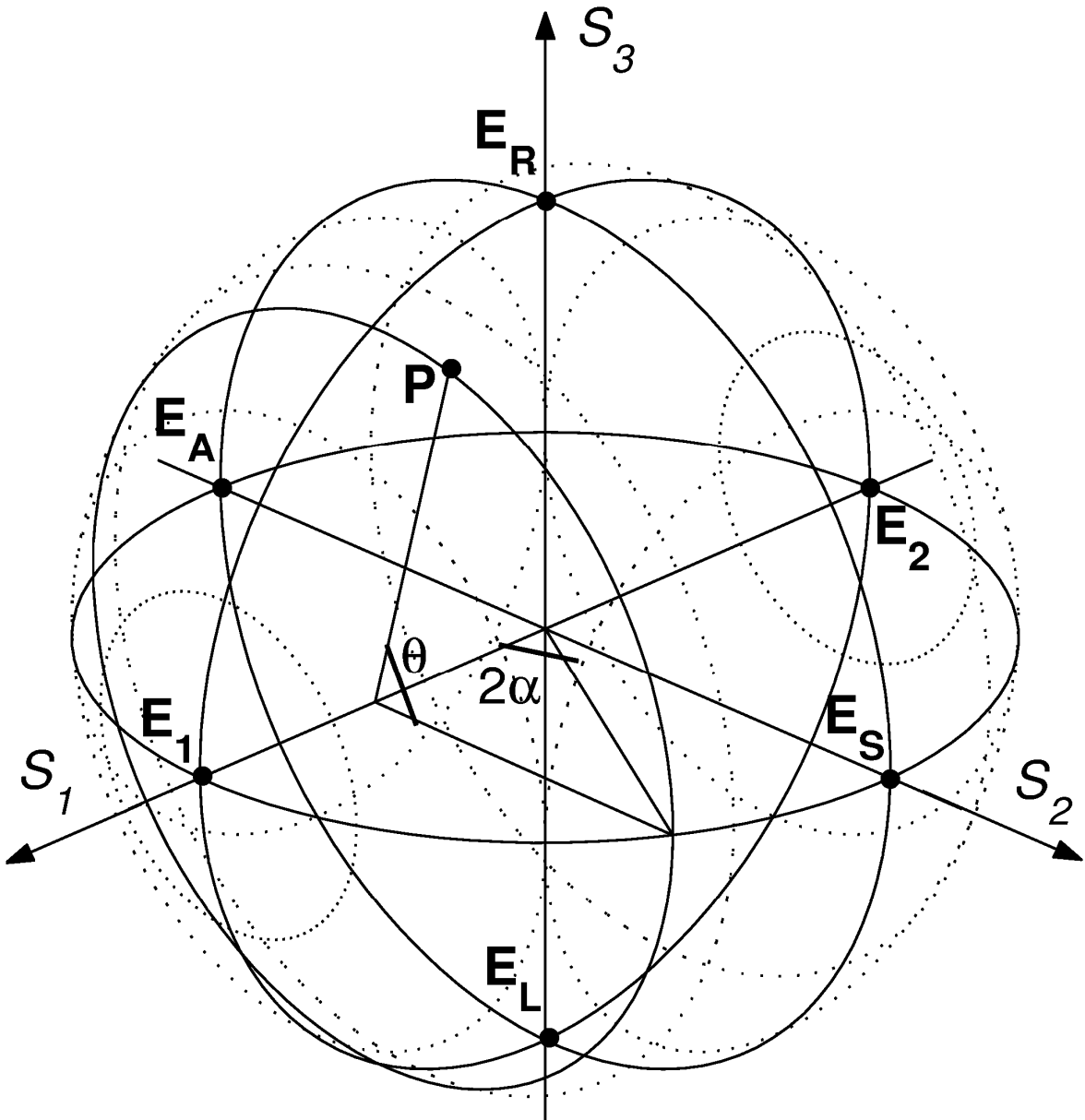

Fig. 2: The Generalized Poincaré sphere.

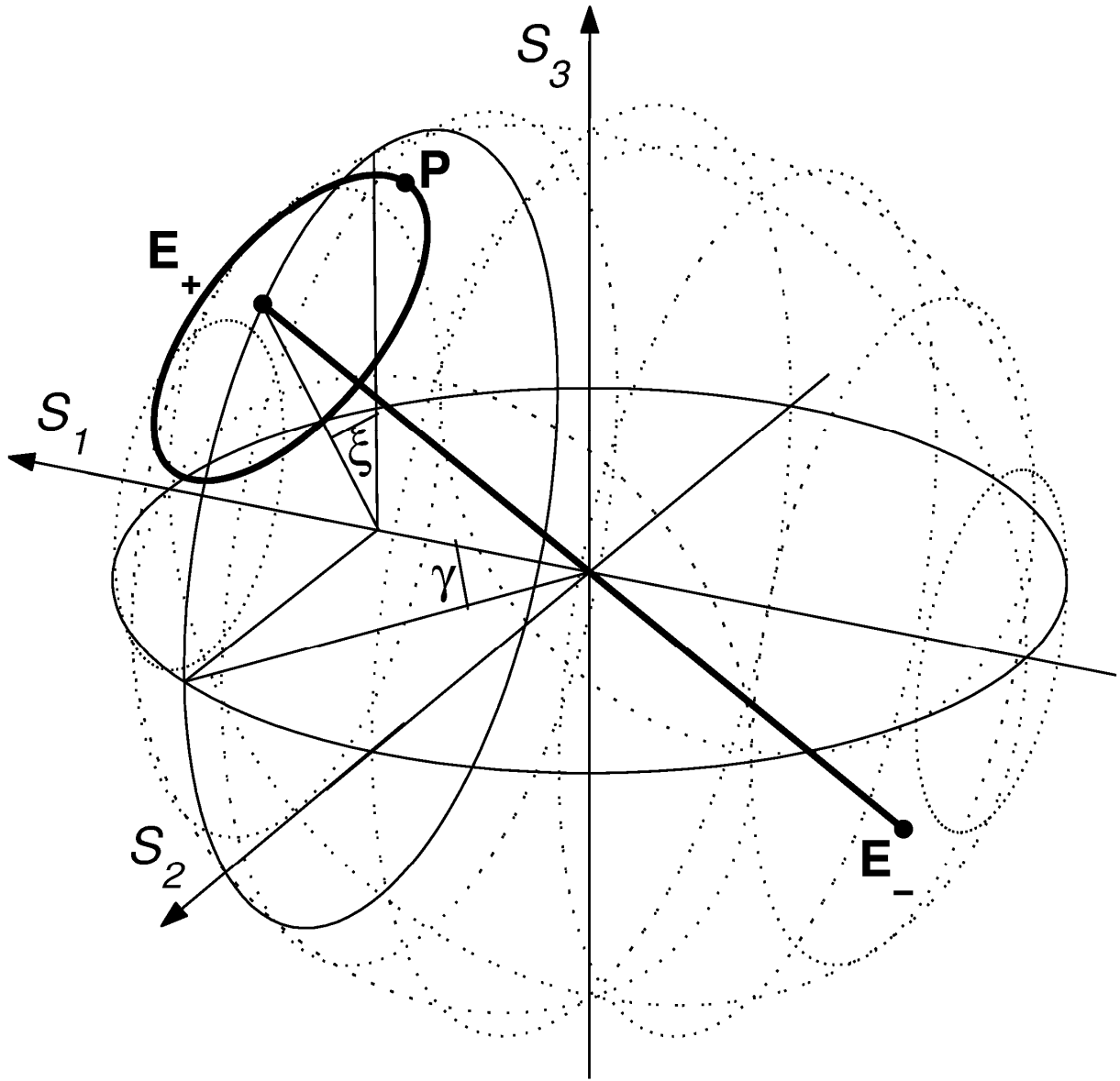

Fig. 3: The eigenstates of $\overline{\mathbf{U}}(\mu L, \xi, \gamma)$.

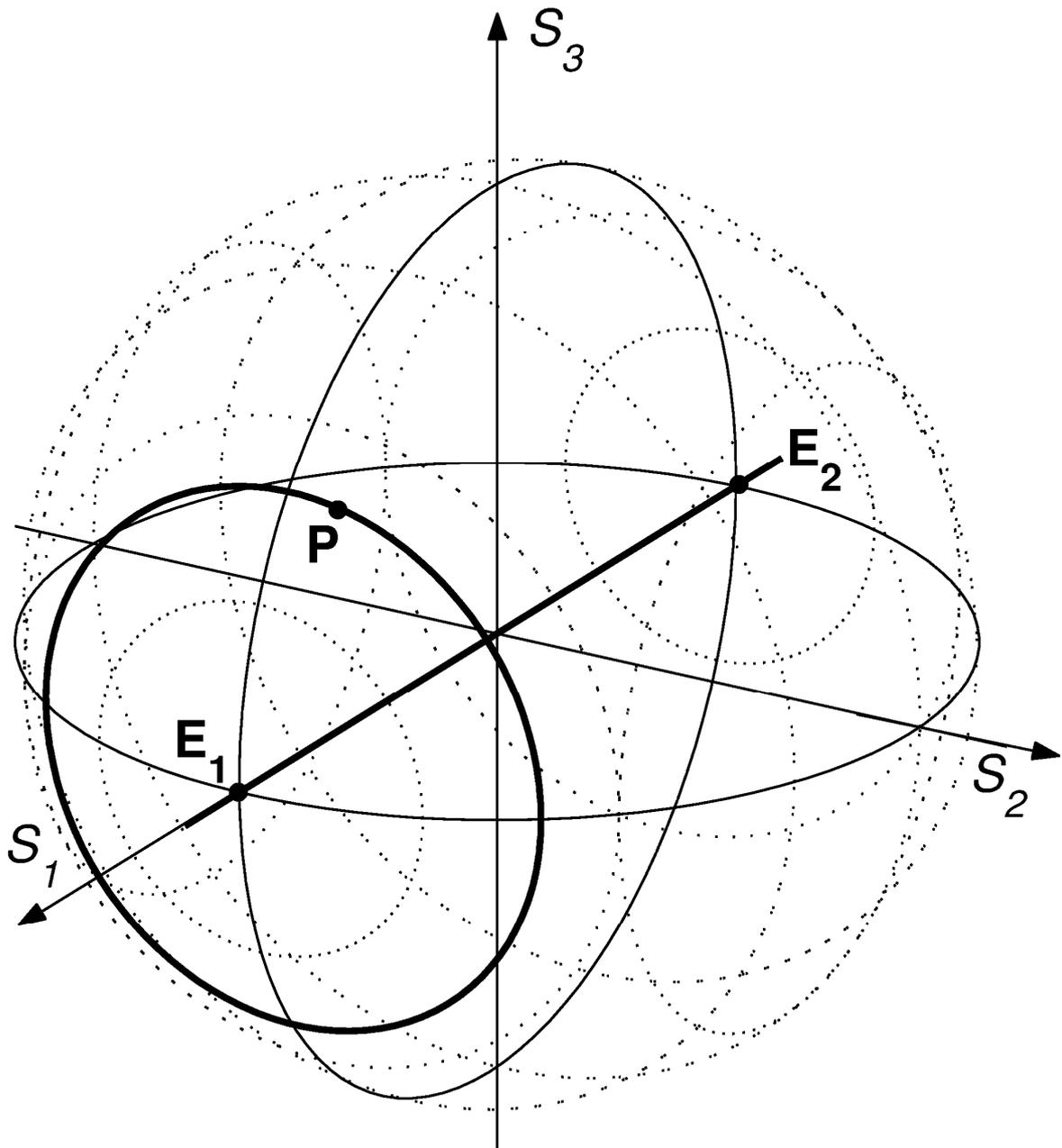

Fig. 4: Action of a phase shifter.

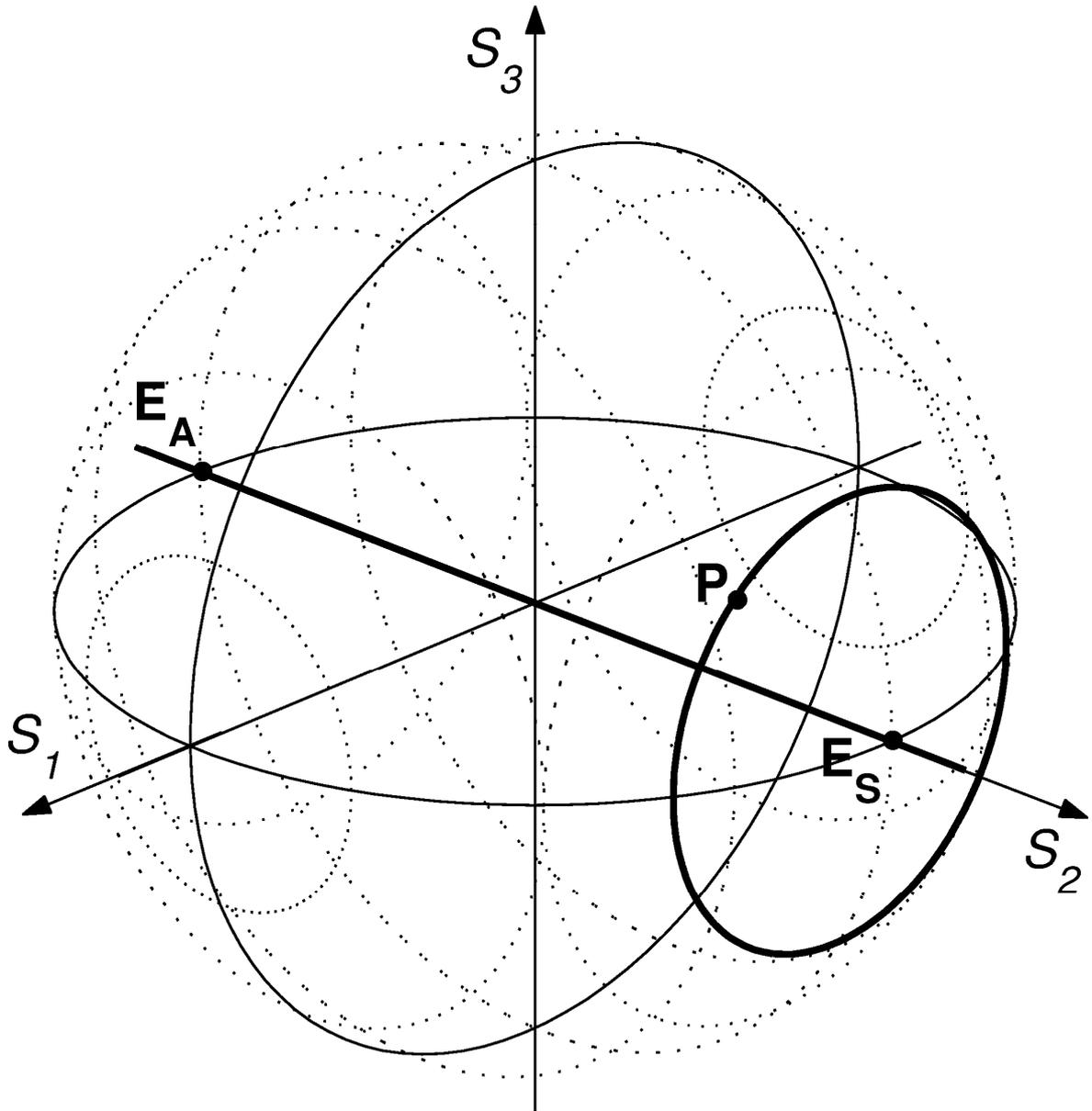

Fig. 5: Action of a synchronous coupler.

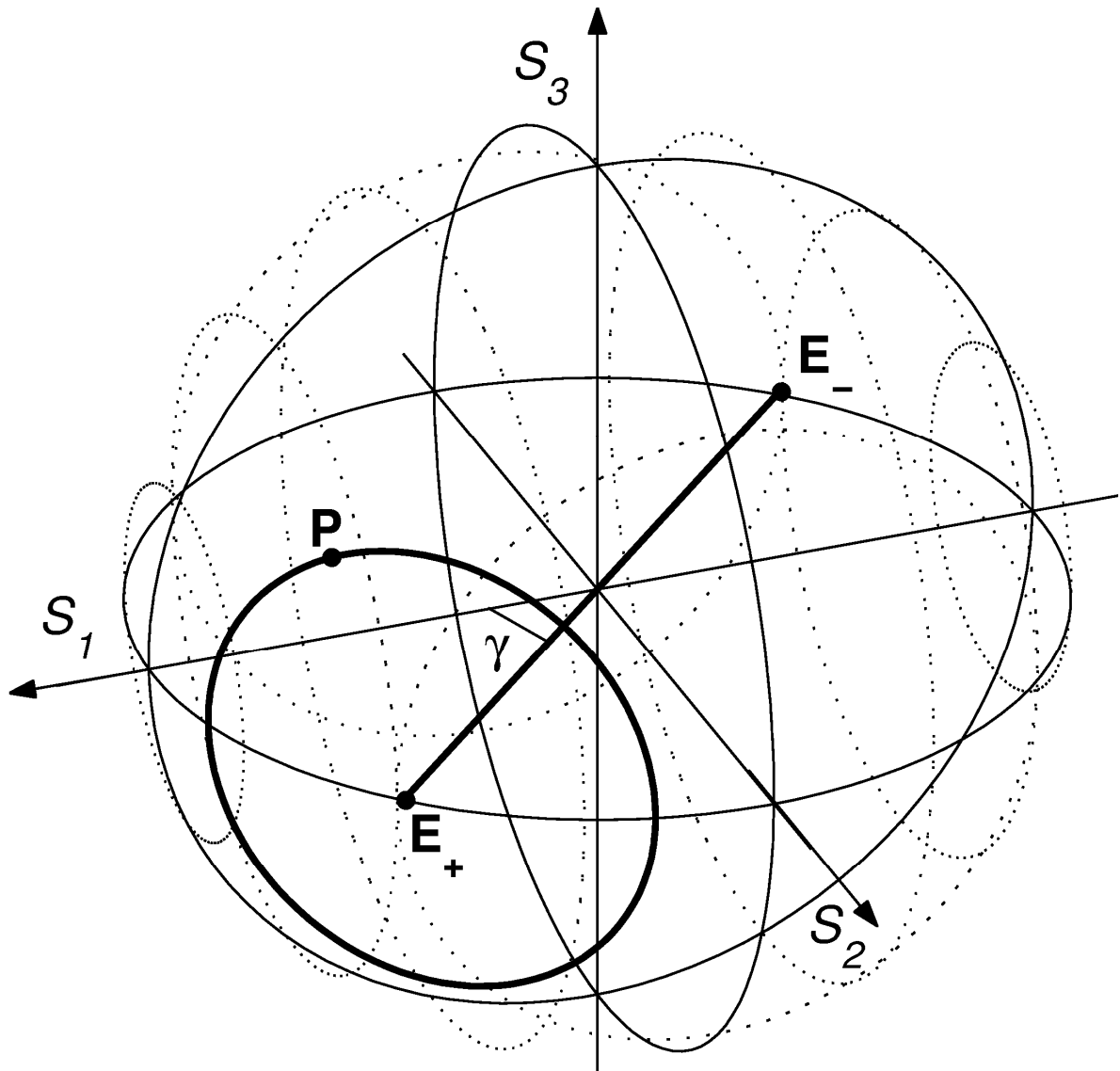

Fig. 6: Action of an asynchronous coupler.

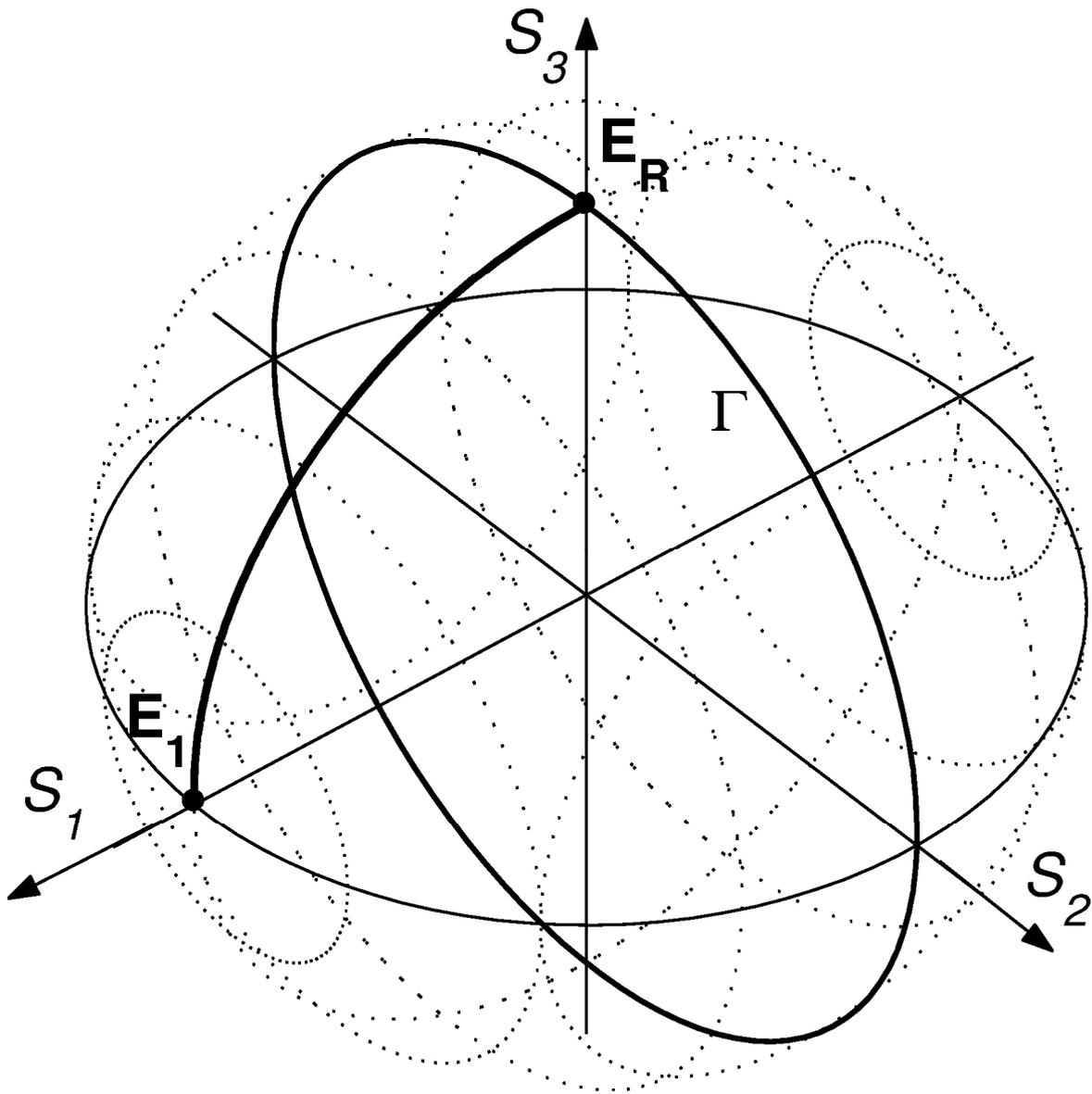

Fig. 7: A 50/50 synchronous coupler.

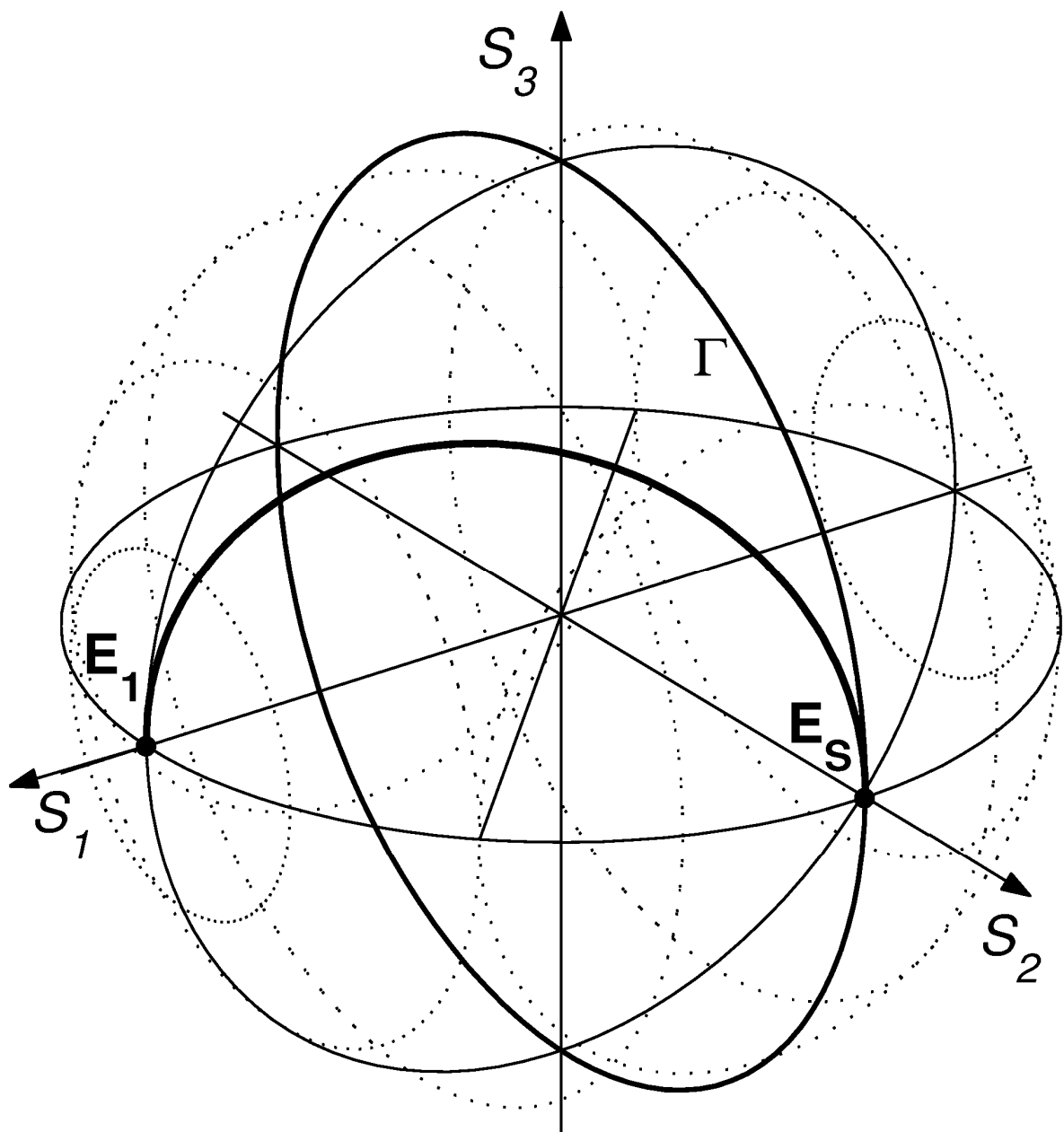

Fig. 8: A 50/50 asynchronous coupler.

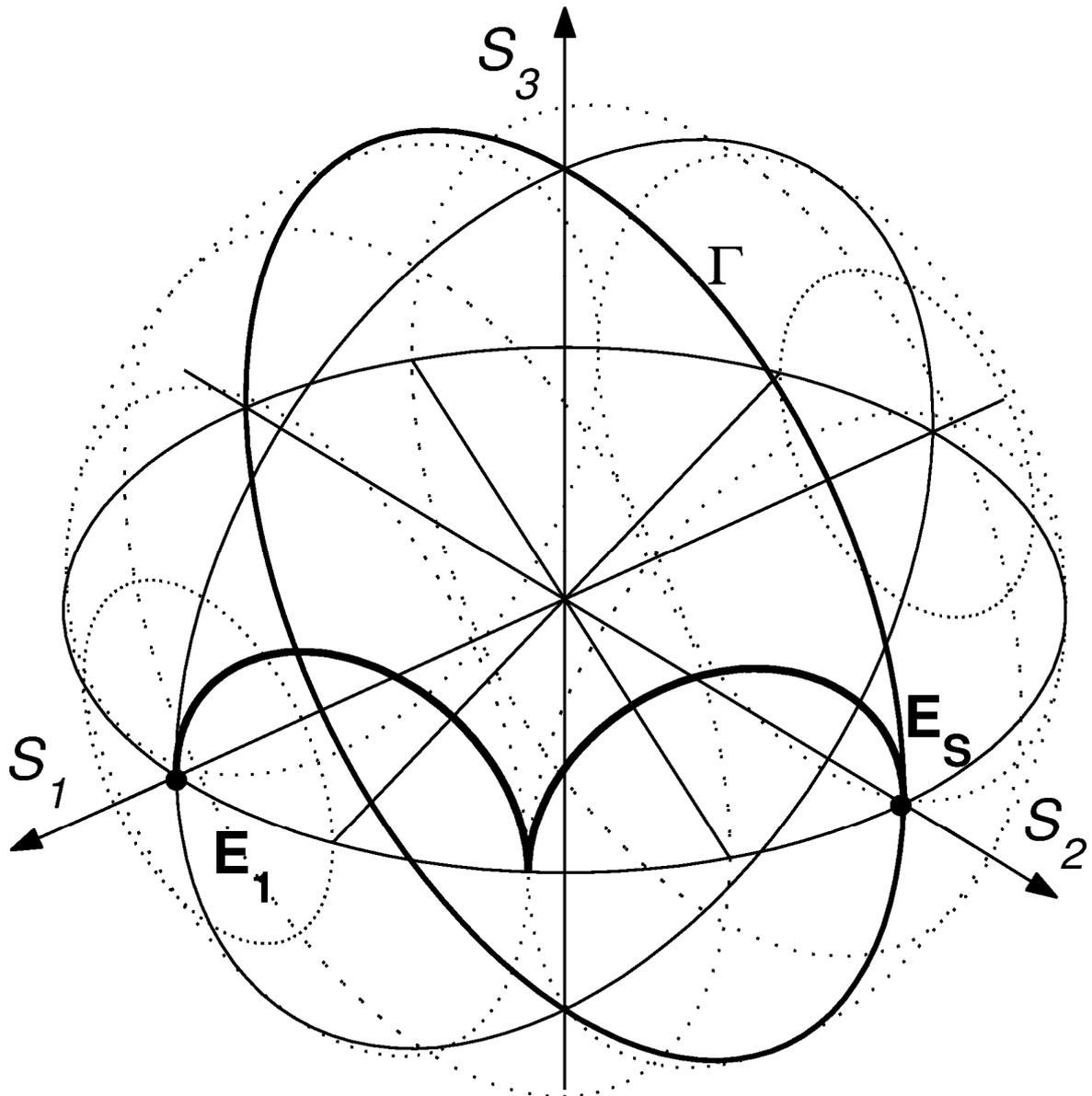

Fig. 9: A 50/50 double asynchronous coupler.

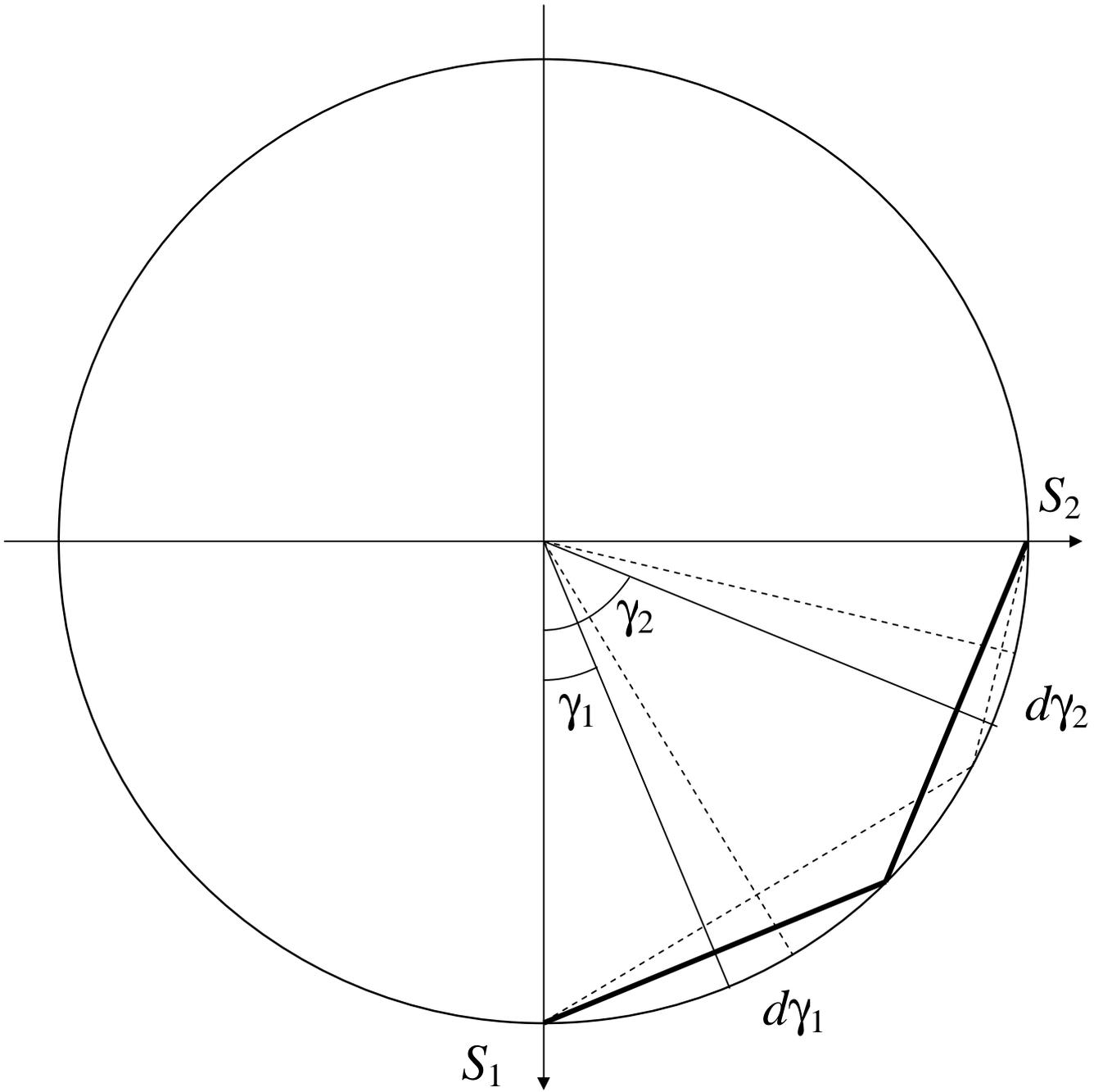

Fig. 10: Insensitivity of the configuration of Fig. 9

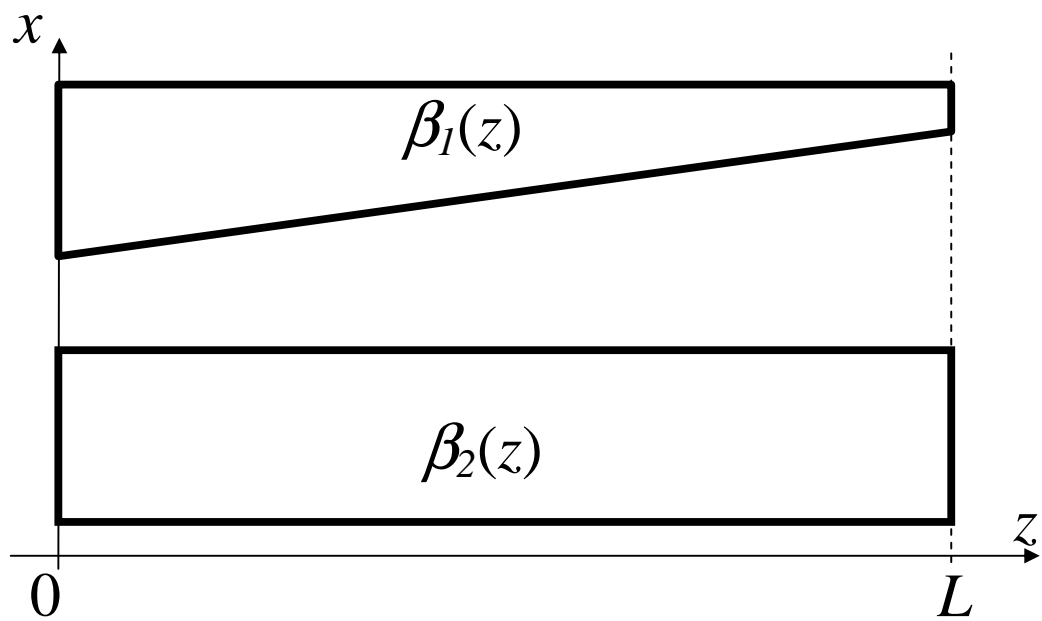

Fig. 11: Schematic of a tapered coupler.

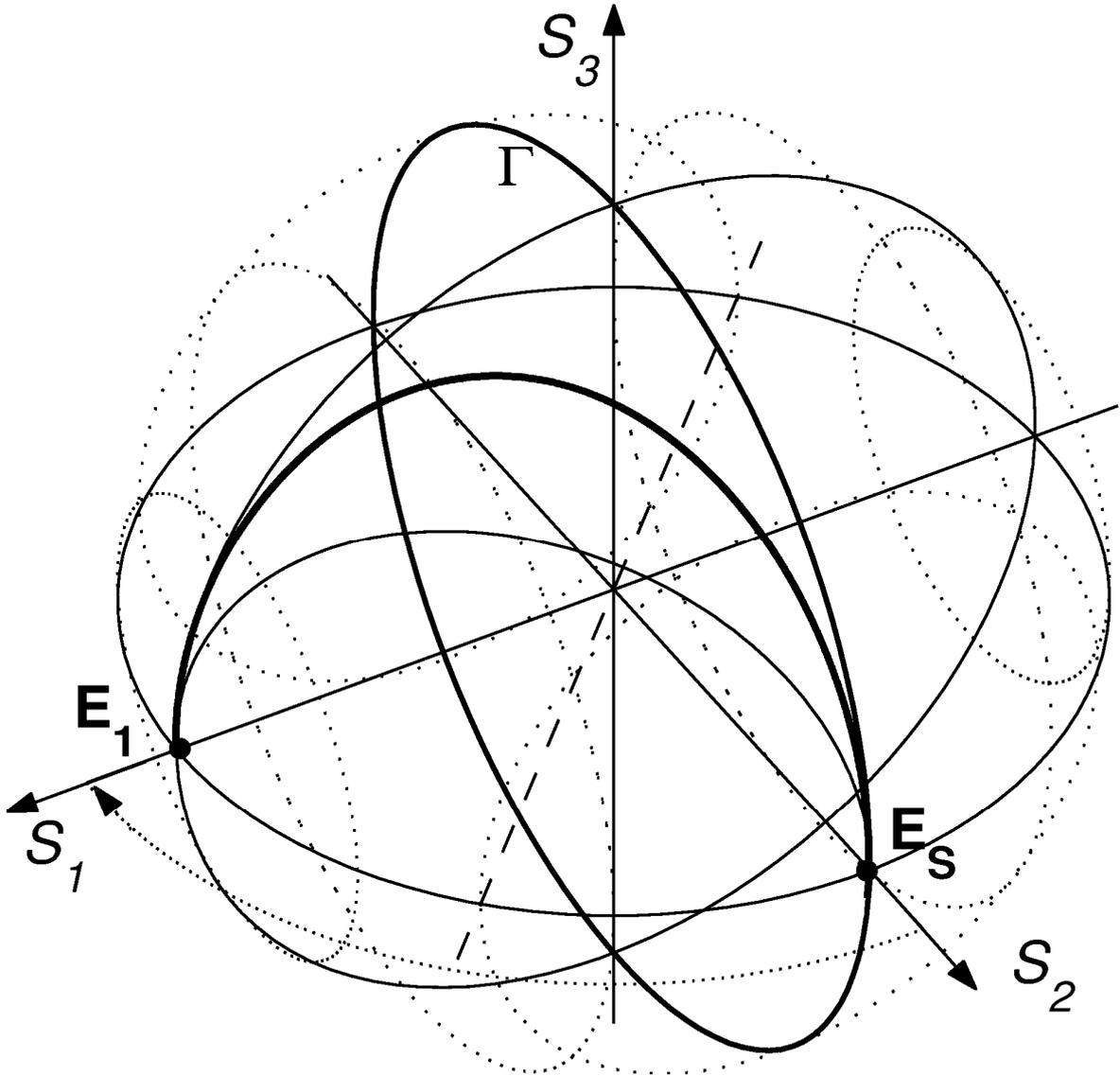

Fig. 12: A 50/50 tapered coupler.

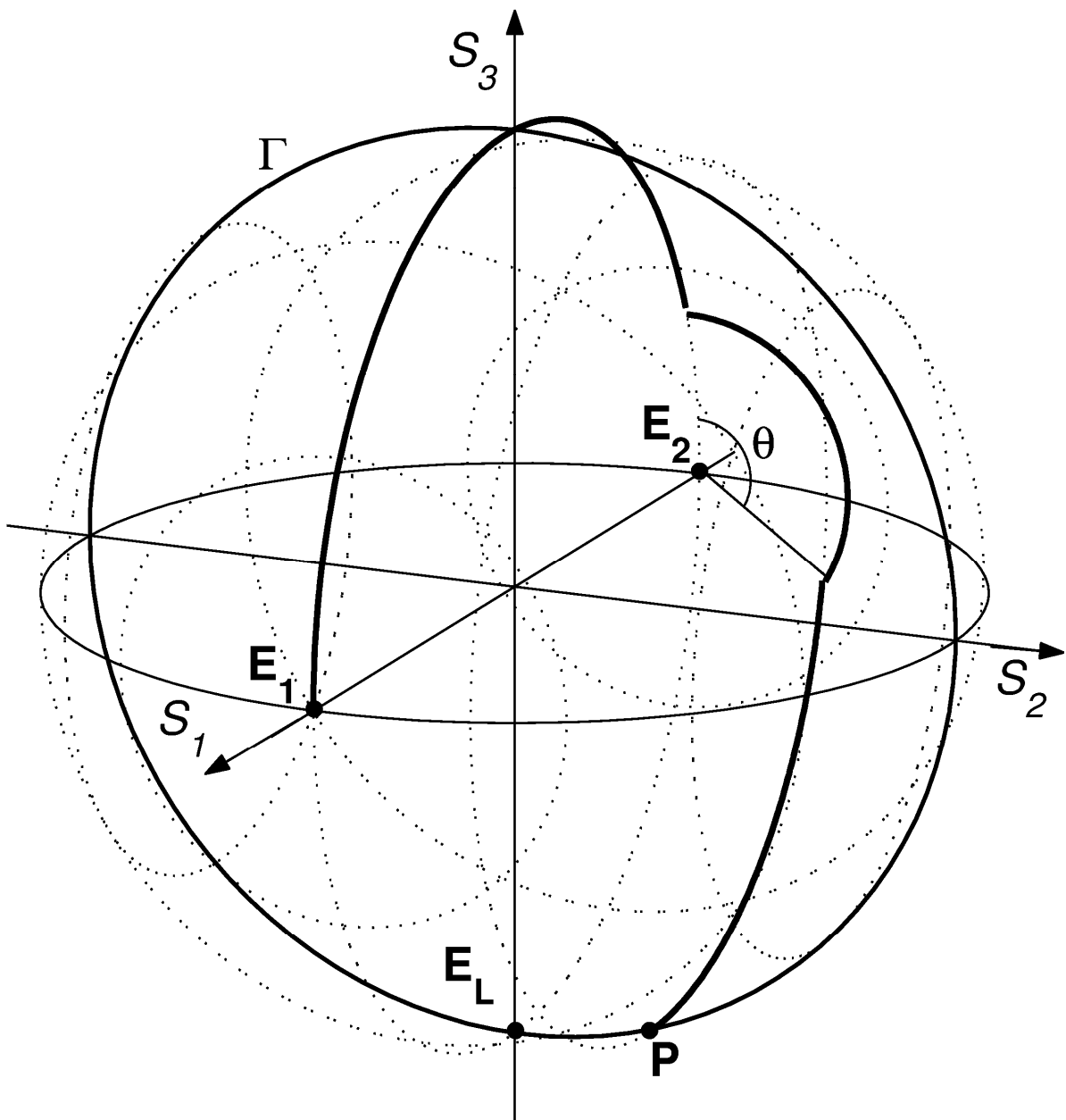

Fig. 13: The 50/50 Mach-Zehnder coupler.